\documentclass[fleqn,10pt]{wlscirep}
\usepackage[utf8]{inputenc}
\usepackage[T1]{fontenc}

\usepackage{relsize}%
\usepackage{amsmath,amssymb}

\title{The evolution of polarization in the legislative branch of government}

\author[1]{Xiaoyan Lu}
\author[1]{Jianxi Gao}
\author[1,*]{Boleslaw K. Szymanski}
\affil[1]{Social and Cognitive Networks Academic Research Center and Department of Computer Science, Rensselaer Polytechnic Institute, Troy NY 12180, USA}

\affil[*]{szymab@rpi.edu}

\begin{abstract}
The polarization of political opinions among members of the U.S. legislative chambers measured by their voting records is greater today than it was thirty years ago. Previous research efforts to find causes of such increase have suggested diverse contributors, like growth of online media, echo chamber effects, media biases, or disinformation propagation. Yet, we lack theoretic tools to understand, quantify, and predict the emergence of high political polarization among voters and their legislators. Here, we analyze millions of roll-call votes cast in the U.S. Congress over the past six decades. Our analysis reveals the critical change of polarization patterns that started at the end of 1980's. In earlier decades, polarization within each Congress tended to decrease with time. In contrast, in the recent decades, the polarization has been likely to grow within each term. To shed light on the reasons for this change, we introduce here a formal model for competitive dynamics to quantify the evolution of polarization patterns in the legislative branch of the U.S. government. Our model represents dynamics of polarization, enabling us to successfully predict the direction of polarization changes in 28 out of 30 U.S. Congresses elected in the past six decades. From the evolution of polarization level as measured by the Rice index, our model extracts a hidden parameter - polarization utility which determines the convergence point of the polarization evolution. The increase in the polarization utility implied by the model strongly correlates with two current trends: growing polarization of voters and increasing influence of election campaign funders. Two largest peaks of the model's polarization utility correlate with significant political or legislative changes happening at the same time.
\end{abstract}

\begin{document}

\flushbottom
\maketitle
% * <john.hammersley@gmail.com> 2015-02-09T12:07:31.197Z:
%
%  Click the title above to edit the author information and abstract
%
\thispagestyle{empty}

\section*{Introduction}

Conflict and consensus play important role in the functioning of a social system. In the context of political competition they manifest themselves as 
polarization of opinions and collaboration to reach consensus on shared national interests~\cite{diamond1990three}. Polarization arises from the politicians' need 
to represent opinions of their voters while collaboration is required to balance the interests of many groups. Numerous previous publications have focused on the role of social conformity~\cite{turner1991social,feldman2003enforcing,klucharev2009reinforcement} in polarization. Among these publications, many hypotheses have been proposed to explain the observed emergence of increased polarization, including social homophily~\cite{mcpherson2001birds}, selective 
exposure~\cite{stroud2010polarization}, social bots~\cite{ferrara2016rise}, echo chambers~\cite{garrett2009echo, scirep18echo}, propagation of low-quality 
information or fake news~\cite{jin2014misinformation, allcott2017social}, as well as the effect of viral news~\cite{lu2018scalable} and social media~\cite{garimella2018polarization}. 
While these models study different aspects of polarization of political views, they share some common assumptions about human social behavior~\cite{dixit2007political}, including the following: (i) individuals iteratively update their views to reach consensus with their neighbors in a social network; (ii) the tolerance of conflicting 
views is limited in social context, so frequent active disagreements usually break of social ties~\cite{Bahulkar2017author}. These assumptions indicate that the loyalty to one's 
group usually leads to the conformity with views of the group's majority~\cite{deutsch1955study}, and such conformity tightens social ties within the group. Therefore,
in our model, we allow the current polarization level to influence its future growth.

We analyze millions of roll-call votes cast in the U.S. Congress~\cite{poole2012voteview} over the past six decades to identify evolution of political polarization patterns. Using the roll-call vote results, we quantify the level of polarization in the legislative branch of government over the last six decades. We assume a social system dominated by two parties. In such a system polarization and collaboration can convert into each other but they maintain their sum constant at 1. A simple model of the dynamics of such conversion~\cite{abrams2011dynamics} can be written as
\begin{equation} \label{eq:model_special}
    \frac{dx}{dt} = y P_{yx}(x,u_x) - x P_{xy}(x,u_x)
\end{equation} 
where $x\in[0,1]$ is the current polarization level, as measured by the real legislative votes, while $u_x$ is a parameter independent of the current polarization level. We call parameter $u_x$ the polarization utility in analogy to the role of gravitational utility in physics. The complementary values denoted as $y = 1 - x$ and $u_y = 1 - u_x$ represent the current collaboration level and the collaboration utility, respectively, while
$P_{yx}(x,u_x)$ is the probability of collaboration converting to polarization per unit of time. For symmetry, $P_{xy}(x,u_x)$ represents the probability of polarization converting to collaboration per unit of time. Studying polarization in the U.S. Congress, we assume the evolution is fully governed by the nonlinear dynamics defined in Eq.~[\ref{eq:conversion}]. However, every two years, the dynamical system moves to a new state determined by the election of members of the next Congress for which this state then becomes the initial state. 

Up to the end of 1980's, the polarization level within each Congress tended to decrease. Since then, however, the polarization has been likely to grow within  two-year term of each Congress, as shown in Fig.~\ref{fig:fig3}A. This phenomenon is represented in our model by the change of the polarization utility, since it determines the polarization level to which the system converges, regardless of its initial state. As illustrated in Fig.~\ref{fig:fig3}B and ~\ref{fig:fig3}C, the non-linear dynamics successfully predict the direction of polarization change in 28 out of 30 U.S. Congresses for the past six decades. The two Congresses that in disagreement with the model predictions have very small variations of polarization, with a weakly-defined polarization direction, making the predication error small.
Moreover, Fig.~\ref{fig:fig3}D suggests that farther away is the initial polarization caused by member replacement from the corresponding equilibrium defined by our model, faster the polarization level changes during the two-year period between elections. 

In our model, the non-linear gain-loss function quantifies the conversion between collaboration and polarization among legislators. The model implies that the polarization level always converges to an equilibrium point, while replacement of members in each Congress caused by election sets the new initial polarization level. We also derive an approximate analytic expression for the equilibrium points, which are defined by the polarization utility and the system's sensitivity to the current polarization. Our model implies that the observed increased polarization in the recent few decades is caused by the growing polarization utility. This conclusion prompts the question about the causes of this growth. We address this question in the Discussion section.
% Place tables after the first paragraph in which they are cited.
 \begin{table}[!ht]
 \centering
\caption{
{\bf The statistics of the roll-call votes from the 85th to the 114th U.S. Congress.}}
\begin{tabular}{|c|c|c|c|c|c|}
\hline
%\multicolumn{4}{|l|}{\bf Heading1} & \multicolumn{4}{|l|}{\bf Heading2}\\ \thickhline
Party & \#Members & \#Bills & \#Votes & \#Votes per member & \#Votes per year\\ 
\hline
Democratic Party & 1498 & 31,879 & 7,368,921 & 4919.17 & 122815.35\\ %\hline
Republican Party & 1395 & 31,879 & 6,275,886 & 4498.84 & 104598.1 \\ 
\hline
\end{tabular}
\begin{flushleft} The bills which received less than 30 votes are not included in the above statistics.%pellentesque metus tortor nec nisl. Ut ornare mauris tellus, vel dapibus arcu suscipit sed.
\end{flushleft}
\label{table1}
\end{table}

\begin{figure}
    \centering
    \includegraphics[width=\textwidth]{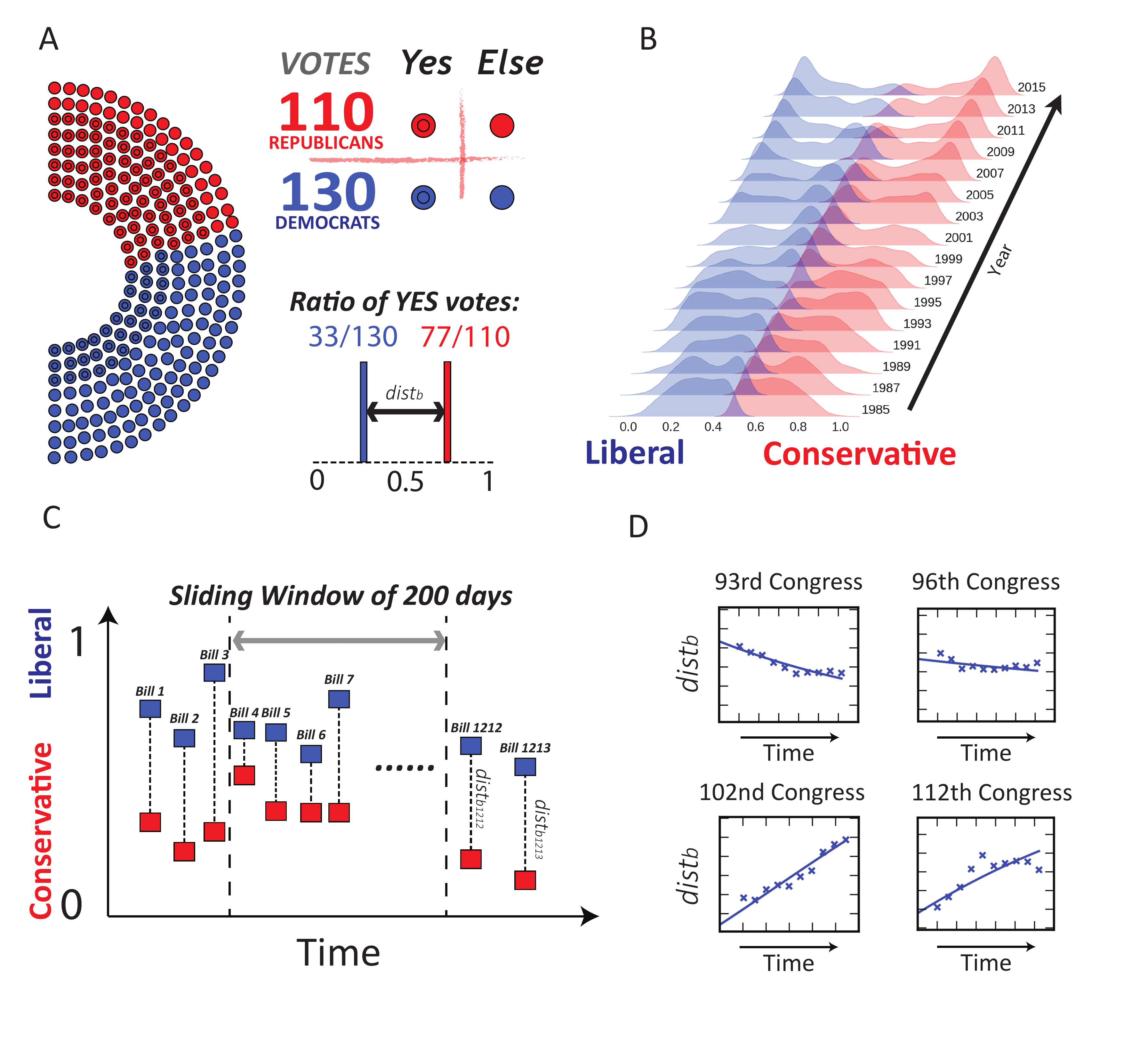}
    \caption{The illustration of the data collections and processing workflow. (A) For each bill voted in the Congress, we measure the mean absolute distance, 
$\textit{dist}_b$, between the votes of the Democratic and Republican Parties cast for the bill $b$; (B) The distribution of the political polarization measured using the roll-call votes within each Congress. Labels at the right corner of each sub-plot identify the first year of each Congress. In the 1980's and 1990's, the polarization levels are generally smaller than the levels in the 2000's and 2010's. After 2001, the two peaks of the voting results at the opposite ends of the political spectrum start to emerge, indicating the growth of the number of bills on which two parties strongly disagree; (C) Regardless of the content of bills, we compute the average distance of bill votes between two parties within each Congress in a sliding window of $200$ days (Eq.~[\ref{eq:average_level}]); (D) The political polarization levels at ten evenly-distributed sampled time points exhibit an evolution of polarization patterns from one type of behavior to another: the polarization level decreases in the 93rd Congress as time from the replacement of members increases, while polarization remains at a relative stable level in the 
$96^{th}$ Congress, and grows in the Congresses with sessions numbered from $102$ to $112$.}
    \label{fig:fig1}
\end{figure}

\section*{Quantifying polarization in the legislative branch}
We analyze millions of roll-call votes\footnote{The dataset is accessible at https://voteview.com/} cast in the U.S. Congress~\cite{poole2012voteview} over the past six decades to identify evolution of political polarization patterns. The statistics of the roll-call votes in Congresses with sessions numbered from 85 to 114 are shown in Tab.~\ref{table1}. This dataset contains approximately 7 million votes in both the Senate and the House of Representatives from a total of 1498 and 1395 legislators from Democratic and Republican Parties, respectively. We adopt the well-known Rice index~\cite{rice1928quantitative} to measure party dissimilarity in legislative voting. The Rice index is defined as the mean absolute distance between the Yes-ratios of Democratic and Republican Parties on the $b^{th}$ bill 
\begin{equation} \label{eq:dist_def}
\textit{dist}_b = \Big| {\mathlarger{\mathbb{E}}}_{\{\mathsmaller{ 1 \leq i \leq N_{Rep}} \}}  \text{Rep}_{ib} - {\mathlarger{\mathbb{E}}}_{\{\mathsmaller{ 1 \leq j \leq N_{Dem}} \}} \text{Dem}_{jb} \Big|
\end{equation}
where $N_{Rep}$ and $N_{Dem}$ denote the numbers of legislators from the Republican and Democratic Parties participating in the vote, while $\text{Rep}_{ib}$ and $\text{Demo}_{jb}$ are the votes cast by Republican $i$ and Democrat $j$ for bill $b$, respectively. $\mathbb{E}$ represents the corresponding average over all legislators in each party. The result of a vote is coded as 1 for Yes and 0 otherwise because the bills pass by majority in both the Senate and the House of Representatives and therefore abstaining is effectively equivalent to opposing bill passage. This procedure is illustrated in Fig.~\ref{fig:fig1}A.

Regardless of the content of bills, we compute the average distance of bill votes between two parties over every $199$ day intervals. Formally, the polarization level at the $t^{th}$h day of the $k^{th}$ Congress is quantified as
\begin{equation} \label{eq:average_level}
x_k(t) = \mathlarger{\mathbb{E}}_{\{\mathsmaller{b: |t_b-t|<100 }\}} \textit{dist}_b 
\end{equation}
where $\{b: |t_b-t|<100 \}$ is the set of bills voted within $199$ days centered at the $^{th}$ day of the $i^{th}$ Congress and $i$ starts at the $100^{th}$ day of each Congress and ends 99 days before the last day of this Congress. Hence, each measurement includes exactly 200 days of voting. This step is illustrated by Fig.~\ref{fig:fig1}C. The averaging reduces the noise of the raw data because the topics of the legislative bills may differ day-by-day in each Congress and there were several periods of times in the past six decades during which very few bills were voted on by the U.S. Congresses. Compared to the approaches~\cite{baldassarri2007dynamics, moody2013portrait, gu2017co} defining the conflicting level between individuals, the well-know Rice index defined in Eq.~[\ref{eq:dist_def}], reflects the general trend of behavioral partisanship while preserving its simplicity by quantifying polarization at party level. More importantly, the Rice index enables us to develop a dynamical equation, Eq. [\ref{eq:modela}], which captures the macroscopic behavior of the evolution of political polarization, regardless of the complex interactions between individual legislators considered in~\cite{moody2013portrait}.

For every Congress and each bill, we measure the distance $\textit{dist}_{b}$ between two parties using Eq.~[\ref{eq:dist_def}]. As $\textit{dist}_b\in[0,1]$, the political preference for bill $b$ is defined as $D_b = 0.5-\frac{\textit{dist}_{b}}{2}$ for the Democratic Party and $R_b = 0.5+\frac{\textit{dist}_{b}}{2}$ for the Republican Party. Then, using the kernel density estimation (KDE)~\cite{silverman2018density}, we evaluate the distribution of distances $D_{b}$ and $R_{b}$, of Democratic and Republican Parties, respectively, from the center of the polarization range. Fig.~\ref{fig:fig1}B shows the distribution of these distances which represent positions of the two parties regarding the bills voted in each Congress. 

\section*{Dynamical model of political polarization}

\begin{figure}
    \centering
    \includegraphics[width=\textwidth]{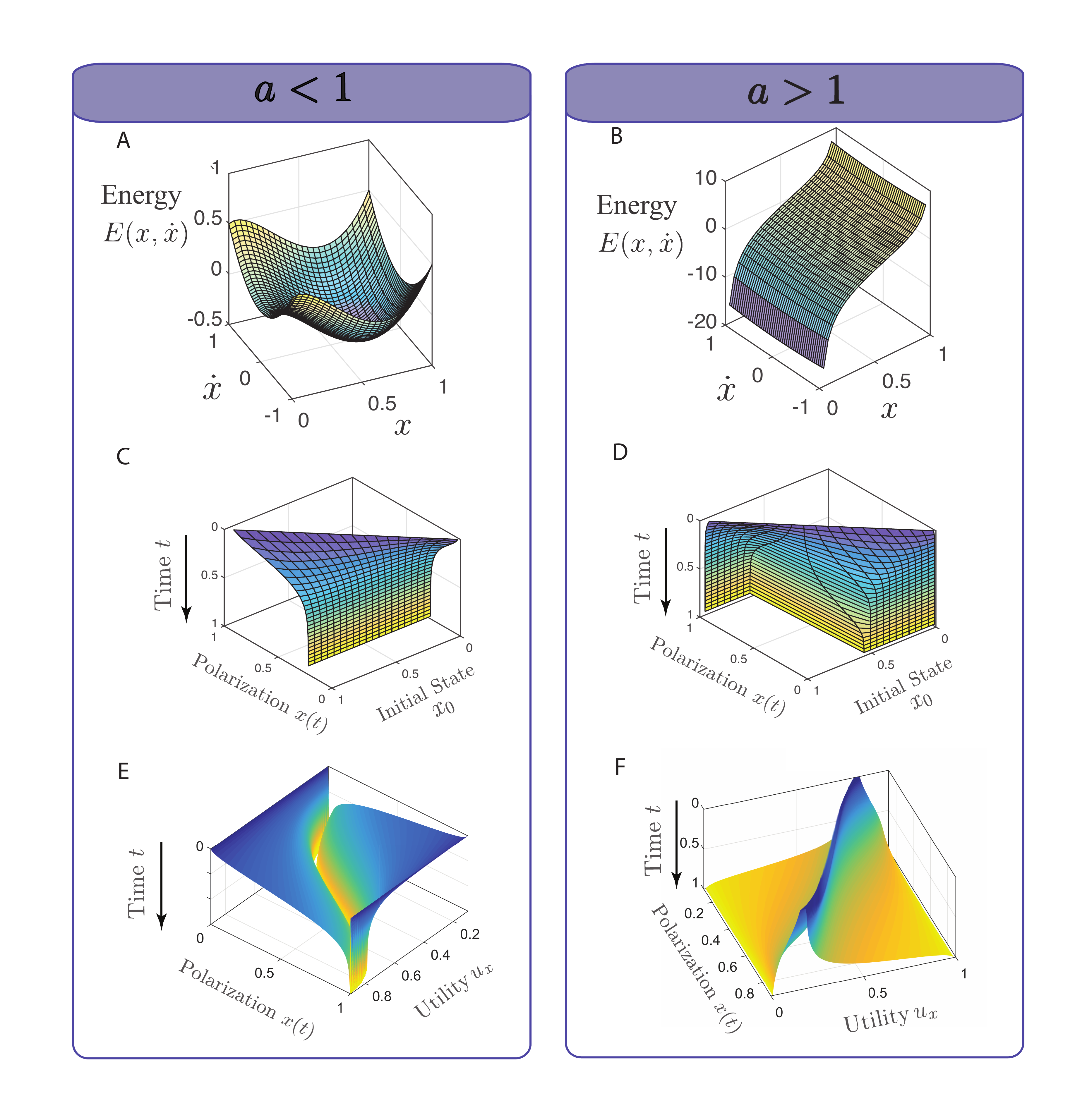}
    \caption{The convergence of the political polarization model in relation to the polarization utility $u_x$ and initial state $x_0$. The time $t$ is normalized in the plots to the range from $0$ to $1$. (A) The total energy of the dynamical system in relation to the polarization $x$ and its first order derivative $\dot{x}$ for $a<1$ (Eq.~[\ref{eq:energy}]); (B) The total energy of the dynamical system in relation to the polarization $x$ and its first order derivative $\dot{x}$ for $a>1$ (Eq.~[\ref{eq:energy}]); (C) For $a<1$ the dynamical system always converges to certain polarization level ($a=0.6$); (D) For $a>1$ the dynamical system either reaches complete consensus or complete polarization depending on the initial state $x_0$ ($a=2.5$). (E) When $a < 1$, the equilibrium points of the dynamical system are stable as the system gets trapped at these equilibrium points as the time approaches infinity; (F) When $a > 1$, the initial states (on the top of the hills) are the tipping points causing the system to converge either to 0 (full polarization) or 1 (full consensus) from its initial state $x_0$.}
    \label{fig:fig2}
\end{figure}

We assume a two-party political system, such as exemplified by the United Kingdom, but applicable also to the U.S. and other countries with a multi-party system dominated by two major parties. We also assume a social system in which polarization and collaboration can convert into each other but preserve their sum at 1.  A simple model of the dynamics of such conversion is given in~\cite{abrams2011dynamics} as
\begin{equation} \label{eq:modela}
    \frac{dx}{dt} = y P_{yx}(x,u_x) - x P_{xy}(x,u_x)
\end{equation} 
where $P_{yx}(x,u_x)$ is the probability of collaboration converting to polarization per unit of time. For symmetry, $P_{xy}(x,u_x)$ represents the probability of polarization converting to collaboration per unit of time. Finally, $x\in[0,1]$ is the current polarization level as measured by the real legislative votes, $u_x$ is the polarization utility parameter, and $y = 1 - x$ and $u_y = 1 - u_x$ are complementary to $x$ and $u_x$, representing the current collaboration level and the collaboration utility ($u_x + u_y = 1$). Following~\cite{abrams2011dynamics}, we assume that $P_{yx}$ has the following simple form
\begin{align} \label{eq:conversion}
P_{yx}(x, u_x) &= c x^a u_x 
\end{align}
which is supported by the normative social influence theory~\cite{deutsch1955study} that postulates that the current polarization level $x$ influences the probability of conversion. To preserve symmetry under the conversion from $y$ to $x$ or vice versa, we define $P_{xy}$ as
\begin{align}
P_{xy}(x,u_x) &= P_{yx}(y,u_y)=cy^au_y=c(1-x)^a(1-u_x)
\end{align}
Similar nonlinear gain-loss equations for the state dynamics have been successfully applied to model various types of polarization, ranging from religious affiliation~\cite{abrams2011dynamics}, to language choice~\cite{abrams2003linguistics} and political affiliation~\cite{frachebourg1996exact}. 
The nonlinear dynamics defined here captures the conversion between polarization and collaboration during different periods. The parameter $u_x$ is independent of the polarization level $x$ which varies as the time changes. It reflects internal hidden polarization level in the legislative branch of government, which is not reflected from the vote results at that moment. The term $x^a$ captures the effect of the current polarization level on the evolution. The speed of evolution is defined by the parameter $c$ in our model.

The total energy~\cite{perko2013differential} in this dynamical system is governed by the following equation.
\begin{equation} \label{eq:energy}
    E(x, \dot{x}) = \frac{1}{2} \dot{x}^2 - c x y \Big( \frac{u_x}{x^{1-a}} - \frac{u_y}{y^{1-a}} \Big)
\end{equation}
which is constant on the solution curves or trajectories of this system, i.e. $E(x, \dot{x}) = C$ for some constant $C$. The total energy function $E(x, \dot{x})$ in relation to the current polarization $x$ and its first order derivative $\dot{x}$ is shown on the Fig.~\ref{fig:fig2}A for $a < 1$ and Fig.~\ref{fig:fig2}B for $a > 1$.
% Quick tutorial found here
% http://www.cds.caltech.edu/~marsden/wiki/uploads/cds140a-08/lecturenotes/MechSystems.pdf
% The reference book includes all the details

Given the model parameters $a$ and $u_x$, the equilibrium points $x^*$ of the dynamical system defined by Eq.~[\ref{eq:modela}] can be derived as
\begin{align} \label{eq:stable_states}
    x^*(u_x,a) \approx \frac{1}{1+e^{2 \frac{1 - 2u_x}{1 - a}}}
\end{align}
The approximation uses the Taylor expansion of $\dot{x} = 0$. Its detailed derivation is presented in the supporting material. The theoretical expression of equilibrium points closely matches the results found by numerical simulations.

The trajectories of the nonlinear dynamical evolution are shown in Fig.~\ref{fig:fig2}C-F where the x-axis represents time $t$ and the y-axis represents the polarization. Each trajectory curve starts from its initial polarization level $x_0$ which is represented by the z-axis. We show the surface of these convergence process for $a < 1$ and the initial polarization in full range from $x_0 = 0$ to $x_0 = 1$, in Fig.~\ref{fig:fig2}E. Fig.~\ref{fig:fig2}D contains similar visualization results with $a > 1$ and initial polarization levels $x_0$ at the tipping points, i.e. the points at the edges in dark blue of Fig.~\ref{fig:fig2}F. The equilibrium points of the dynamical model are stable when $a<1$. In such case, the dynamical system of Eq.~[\ref{eq:modela}] always converges to the $x^*\in[0,1]$ after the sufficient period of evolution; Fig.~\ref{fig:fig2}C and ~\ref{fig:fig2}E show example with $a=0.6$. In contrast, when $a>1$, the final state of the dynamical system depends on the initial state and the position of the tipping point; Fig.~\ref{fig:fig2}D and ~\ref{fig:fig2}F show example with $a=2.5$. The dynamical system will eventually converge to either $x=0$ or $x=1$.
%, depending on the initial state $x_0$ and the tipping points.

%A stable system always converges to certain polarization level ($a=0.6$ as in Fig.~\ref{fig:fig2}(C) and ~\ref{fig:fig2}(E)), yet when $a>1$, the dynamical system either reaches complete consensus, or becomes fully polarized, depending on the initial state and perturbation ($a=2.5$ as in Fig.~\ref{fig:fig2}(D) and ~\ref{fig:fig2}(F)). 
In the context of political polarization, the case with $a<1$ corresponds to a healthy political system which maintains the polarization level within a certain range. However, the case with $a>1$, corresponds to a system which switches easily between fully polarized equilibrium point and complete consensus convergence point. In the U.S. political system, a change of the system initial state happens periodically every two years as a result of election or reelection of the legislators. The system near the tipping point at the end of one Congress is vulnerable to extreme state switching even under a small change of the initial 
polarization for the next Congress.

We simulate the non-linear dynamics defined by Eq.~[\ref{eq:modela}] until the polarization level $x$ converges. When $a<1$, the equilibrium point is exactly the final polarization level of convergence. 
When $a>1$ the only equilibrium points are $x=1$ that corresponds to the full polarization, and $x=0$ that represents the opposite case of full consensus. Moreover some of the initial states are unstable; they are the tipping points from which the dynamical system nondeterministically converges to either of the two stable equilibrium points.
%For $a>1$ the equilibrium points, except the ones at $x=1$ and $x=0$, are unstable; they are the tipping points from which the dynamical system can non-deterministically converge to either of the two stable equilibrium points, one representing the consensus state and the other its opposite, full polarization state. 
We grid-search for them with different values of $u_x$. In each iteration, we increase the value of $x$ by a small $\Delta x$. If the final convergence states have changed from one state to another due to the increment of $\Delta x$, then the current $x$ is identified as the tipping point.

\begin{figure}
    \centering
    \includegraphics[width=\textwidth]{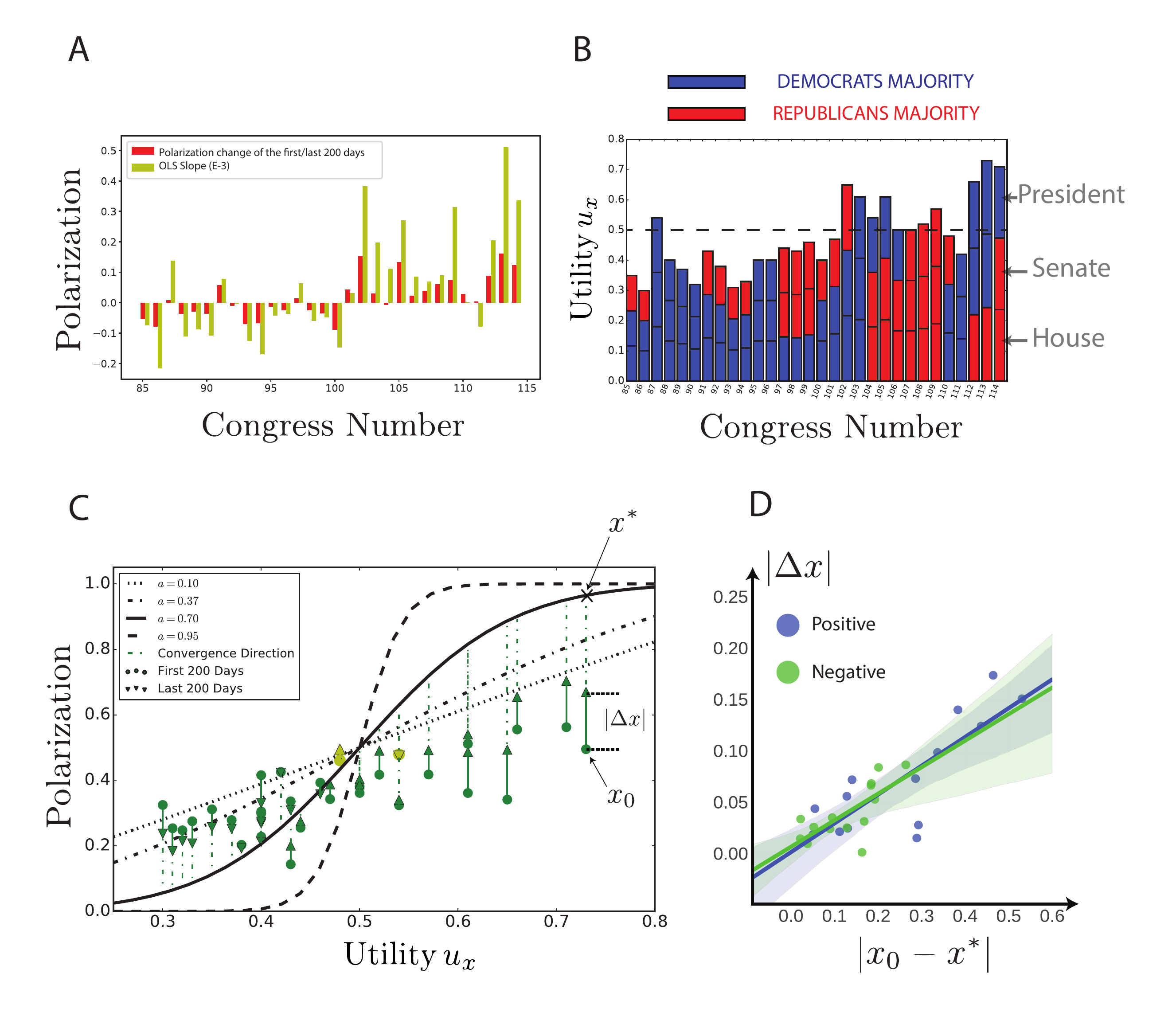}
    \caption{The evolution of the political polarization in the U.S. Congress with subsequent sessions numbered from 85 to 114.
    (A) Change of polarization within each U.S. Congress based on data and the optimal slope parameters estimated by the Ordinary Least Squares (OLS) linear regression given the ten evenly-distributed in time points of polarization sampling for each Congress. The polarization level is likely to decrease within each Congress after the member replacement in the 1970's and 1980's, however, after the $101^{th}$ Congress, which started in 1989, the polarization level is likely to increase instead; (B) The estimated values of the polarization utility $u_x$ are generally increasing, while the periods of sharp growth are often associated with the change of majorities in the Senate and the House of Representatives; (C) When the initial polarization level (green cycles) is smaller than the stable polarization level predicted by our model (solid black curve), we observe an increase of polarization within one Congress. The direction of such change in $28$ out of all $30$ Congresses are explained by the model (green arrows), while the two Congresses in disagreement with prediction have minimal variations of the polarization level (yellow arrow markers) which indicates weakly formed polarization direction; (D) When the initial polarization level $x_0$ of a Congress is farther away from the corresponding equilibrium point, the absolute change of polarization during the two-year term, i.e. $|\Delta x|$, of this Congress is usually higher than of Congresses in which the initial polarization levels are closer to the equilibrium points. The color of the scatter points indicates the sign of $\Delta x$.}
    \label{fig:fig3}
\end{figure}

\section*{Evolution of political polarization patterns}
We fit the political polarization model defined by Eq.~\ref{eq:modela} to the data points $x_i(t)$ which represent ten evenly-distributed in time sampled polarization levels at $t_i=t_{i,1},t_{i,2},\dots,t_{i,10}$ of the $^{th}$ Congress. The system is assumed to have the universal $a$ and $c$ values at all times, but each Congress $i$ has its own parameter $u_x$ defined as the polarization utility, and the particular initial state, $x_{i,0}$, which is defined by the member replacements caused by the most recent election.

The inference procedure estimates the values of universal parameters at $a=0.7$, $c=0.37$ while the set of values for $u_x$, $x_0$ is shown in Fig.~\ref{fig:fig3}B-C. 
In Fig.~\ref{fig:fig3}B, we illustrate the estimated value $u_{i,x}$ of the $i^{th}$ Congress. In general, this value increases as the Congress session numbers grow from 85 to 114. The black solid curve in Fig.~\ref{fig:fig3}C shows the equilibrium points for different $u_x$ values with global parameter $a$ estimated from the voting data. To illustrate the impact of this parameter on values of the equilibrium points, the dotted lines represent these values for $a=0.1$, $a=0.37$ and $a=0.95$, respectively. This figure also illustrates that most of the Congresses last too short to allow the equilibrium point to be reached before the election or reelection of members defines new starting point of the system for the next Congress.

Since the estimate $a=0.7$ is smaller than 1, the dynamical system has a unique equilibrium point for any given $u_x$. As shown in Fig.~\ref{fig:fig3}C, when the polarization level $x$ is larger than the stable equilibrium point $x^*$, it decreases to approach this point. This explains why the polarization level decreases in most of the Congresses in the 1970's and 1980's. After the member replacement caused by the election, the initial polarization levels $x_0$'s during this period were usually higher than the corresponding stable equilibrium points of the dynamical system, see Fig.~\ref{fig:fig3}C. Therefore, the polarization level decreases within the two-year term Congress. In contrast, when the equilibrium point $x^*$ is larger than the initial polarization $x_0$, the polarization $x$ increases over time to approach the equilibrium point. This explains why the polarization level increases in most Congresses after the 1980's. Hence, the observed polarization patterns are fully recreated by the dynamics of our model.

\begin{table} [!ht]
\centering
\caption{The estimated values of the polarization utility (P-utility) $u_x$ in each Congress. The 14 Congresses with the presidential election held in the preceding year increase $u_x$ on average by 11.3\% since the corresponding previous Congress while the 15 Congresses with midterm elections (as the President passes half of his term at the time of the election), decreased the polarization utility $u_x$ by -1.5\% on average. Moreover, in the first three decades, the average polarization utility grew slowly by 0.056, so 18.1\% on average in Presidential election Congresses. All this growth was gained in four Presidential elections in which the newly elected President and his predecessor belonged to different parties; each of these elections contributed growth of 0.115. so 36.4\%, on average. In contrast, the polarization utility decreased by -0.043 or -9.3\% in midterm election Congresses, raising only 14.3\% over 30 years. 
In the latest three decades, the polarization utility grew in both types of Congresses with similar average rates, of  0.023 or 4.6\% for midterm election Congresses, and 0.21 or 7.3\% for Presidential election Congresses. From the $100^{th}$ Congress to $114^{th}$ Congress the polarization grew 77.5\%, so five times higher than in the earlier period. Finally 6 of 14 Presidential election Congresses started with polarization at least 50\% while only one of 15 midterm election congresses achieved such high polarization} \label{tab:utility}
\begin{tabular}{ l  c  c  c  l  c  c  c }
%\hline
\multicolumn{4}{ c }{\bf Midterm Election Congresses} & \multicolumn{4}{ c }{\bf Presidential Election Congresses} \\
\hline
\textbf{Number} & \textbf{P-utility} & \textbf{Change} & \textbf{Percentage}
& \textbf{Number} & \textbf{P-utility} & \textbf{Change} & \textbf{Percentage}\\
\midrule
86$^{th}$ & 0.3 & -0.05 & -14.3\%	& {\bf 87}$^{th,*}$ & {\bf 0.54} & 0.24 & 80.0\% \\
88$^{th}$ & 0.4 & -0.14 & -25.9\%       &  89$^{th}$ & 0.37 & -0.03 & -7.5\% \\
90$^{th}$ & 0.32 & -0.05 & -13.5\%      &  91$^{st}$ & 0.43 & 0.11 & 34.4\% \\
92$^{nd}$ & 0.38 & -0.05 & -11.6\%      &  93${rd}$ & 0.31 & -0.07 & -18.4\% \\
94$^{th}$ & 0.33 & 0.02 & 6.5\%         &  95${th}$ & 0.4 & 0.07 & 21.2\% \\
95$^{th}$ & 0.4 & 0.00 & 0.00\%         &  97${th}$ & 0.44 & 0.04 & 10.0\% \\
98$^{th}$ & 0.43 & -0.01 & -2.3\%       &  99$^{th}$ & 0.46 & 0.03 & 7.0\% \\
100$^{th}$ & 0.4 & -0.06 & -13.0\%      & 101$^{st}$ & 0.47 & 0.07 & 17.5\% \\
102$^{nd}$ & 0.65 & 0.18 & 38.3\%       & 103$^{rd,*}$ & {\bf 0.61} & -0.04 & -6.2\% \\
104$^{th}$ & 0.54 & -0.07 & -11.5\%     & 105$^{th,*}$ & {\bf 0.61} & 0.07 & 13.0\% \\
106$^{th}$ & 0.5 & -0.11 & -18.0\%      & 107$^{th,*}$ & {\bf 0.5} & 0.00 & 0.0\% \\
108$^{th}$ & 0.52 & 0.02 & 4.0\%        & 109$^{th,*}$ & {\bf 0.57} & 0.05 & 9.6\% \\
110$^{th}$ & 0.48 & -0.09 & -15.8\%     & 111$^{th}$ & 0.42 & -0.06 & -12.5\% \\
112$^{th,*}$ & {\bf 0.66}&0.24&57.1\%  & 113$^{th,*}$ & {\bf 0.73} & 0.07 & 10.6\% \\
114$^{th}$ & 0.71 & -0.02 & -2.74\%     & & & & \\
%\hline
\midrule
86th-100th & 0.370 & -0.043 & -9.3\% & & 0.421 & 0.056 & 18.1\% \\
101st-114th & 0.580 & 0.021 & 7.3\% & & 0.559 & 0.023 & 4.6\% \\
All & 0.468 & -0.013 & -1.5\% & & 0.481 & 0.039 & 11.3\% \\
\hline
\multicolumn{4}{ r }{\bf 4  positive, 10 negative, 1 *} & \multicolumn{4}{ r }{\bf 9 positive, 4 negative, 6 *} \\
\end{tabular}
* denotes growth over 50\%, marked also by bold font in the P-utility (polarization utility) column
\end{table}

It is worth noting that the initial polarization levels of the Congresses in the 1990's are actually not significantly higher than those observed in the previous Congresses. However, the polarization utility $u_x$ has become larger and in the later decades exceeded $0.5$ as Fig.~\ref{fig:fig3}B and Table~\ref{tab:utility} indicate. Consequently, the polarization levels at the end of Congresses have significantly increased. This explains the rapid growth of polarization in the later Congresses. The sudden growth of polarization utility in the $101^{st}$ and $102^{nd}$ Congresses (1989-1993), revealed by our model is in agreement with~\cite{moody2013portrait} which describes a dramatic change of polarization that started during the Clinton's term (1993-1994), and solidified during the $104^{th}$ Congress (1995-1996). Thus, the growth of polarization utility came right before the growth of polarization because legislators need time to adjust voting to the increased polarization which has been reflected by the polarization utility. As seen in Fig.~\ref{fig:fig3}D, the absolute change of polarization $|\Delta x|$ grows as the distance between the initial state $x_0$ and equilibrium point $x^*$ increases. If the initial polarization level of a Congress is far away from the its final point of convergence, then the rapid change of polarization would be expected within the two-year term. The direction of such change in $28$ out of all $30$ Congresses are explained by the model, while the two Congresses in disagreement with the model prediction have very small variations of polarization level as shown by the yellow markers in Fig.~\ref{fig:fig3}C. 

Another sign of changed polarization patterns is the number of Congresses in which initial polarization utility is at least 50\%, making it equal to or stronger than then the collaboration utility as defined in Eq.~[\ref{eq:model_special}]. Only one Congress among 15 in the first three decades reached this level, while it
was achieved by 11 out of 14 Congresses in the last three decades.

In summary, our model explains the observed polarization patterns. 
In the 1970's and 1980's, the initial polarization $x_0$ is generally larger than the stable polarization $x^*$, so we observe a decrease of polarization within each of two-year term Congress. In other words, the legislators  gradually agree more and more with the members of the opposite party than initially, while they held more conflicting views at the very beginning of each Congress session. After the $101^{st}$ Congress (1989-1991), the initial polarization $x_0$ is generally smaller than the stable polarization level $x^*$. Therefore, during the corresponding Congresses, the polarization $x$ increases to approach the corresponding equilibrium point defined by the given $u_x$. This trend matches the transition observed during the $103^{rd}$ and $104^{th}$ Congresses~\cite{moody2013portrait}, when the 
moderate members of each party joined their majority-party coalitions, leaving the middle ground deserted.

\section*{Discussion}
The dynamical model sheds some light on the causes of increased polarization in recent decades. As seen in Fig~\ref{fig:fig3}C, when $u_x$ is small, the 
resulting stable polarization level decreases as the value of $a$ increases, in response to the decreased probability of conversion from collaboration to 
polarization (cf.Eq.~[\ref{eq:modela}]). In this case, the dynamical system is less sensitive to the current polarization level $x$ because the term $x^a$ 
in Eq.~[\ref{eq:modela}] becomes smaller as $a$ grows. Therefore, when $u_x$ is small, a large value of $a$ decreases the polarization level. However, when 
the value of $a$ exceeds 1, some initial states become tipping points, from which the system evolves undeterministically towards one of the two possible 
extreme equilibrium points, one of which,  $x=1$, corresponds to the full polarization, while the other, $x=0$, represents the full consensus. Below the 
tipping point, the system converges to one of these two points, and above, it evolves towards the other. 
%Morover, the system in state near such tipping point, may in the presence of externally induced state change, move to the opposite side of the typing point swithing from one extereme equilibrium point to the other. 
In the U.S. political system, a change of the initial system state happens periodically every two years when the legislators are elected or reelected. The 
system in the neighborhood of a tipping point is  vulnerable to even small change in initial polarization that may switch the system convergemce point
from one extreme equilibrium point to another. Such abrupt and radical equilibrium point switching is absent when $a$ is smaller than 1.

According to our model, we witness a growth of the polarization utility, $u_x$, over the recent decades. 
The question arises what are the causes of this increase. To answer this question, we start by observing that a politician needs to be elected to become legislators and repeatadly reelected to continue in this role. Thus, the polarization utility for them lies in its ability to bring votes. This can be accomplished either directly, by representing voters' opinions, or by gaining resources for election campaign funding. In the case of direct support, the polarization of voters has been rising in recent decades for such reasons as the echo chambers effect~\cite{garrett2009echo, scirep18echo} and the growth of new, often strongly biased social media~\cite{garimella2018polarization} or spread of misinformation~\cite{jin2014misinformation, allcott2017social}. Voters increasing polarization raises the polarization utility, which is indirectly reflected as the phenomenon that legislators align their voting with positions of their electorates. The indirect support is gaining in importance because of escalating costs of political campaigns fueled by the growing numbers of effective advertising channels and raising costs of advertising~\cite{ansolabehere2001corruption}. 

To corroborate this conclusion, we identified two largest jumps in polarization utility resulting from election of Congress members (see 
Table~\ref{tab:utility}). The first jump of 0.24 happened in 1960 so it coincides with the start of civil right movement, increasing the U.S. involvement in the Vietnam War, and generational changes in politics. In~\cite{baldassarri2007dynamics}, the author observe that such ``takeoff situations,'' significantly increase polarization in the network structures of political connections. The second jump happened in year 2010, when the Supreme Court approved Super PACs which are allowed to collect unlimited contributions from many sources and to advocate for or against political candidates. 
%The impact of Super PACs might be tamed by a legislative action, like requiring that corporation contributions need to be approved by the majority of the corporation stock owners, and not by the corporation management, as currently is practiced.  However, 
Taming the causes of increased polarization is difficult. For example, requireing biased social media to provide time to advovates of the opposing opinions 
was shown to be counter productive~\cite{Bail9216author}.

The model introduced here implies that high polarization of voters makes the polarization utility higher to legislators, resulting in higher polarization in the legislative chambers of government. The question arises under what conditions the polarized politicians may in turn influence their electorates to become polarized even more. Finding these conditions is important because should such feedback loop arise, it might destabilize democracy. In our future research, we will 
attempt to address this question by developing a quantitative model of polarization dynamics of voters, to large extent shaped by the economic factors~\cite{gelman2011dynamics} and public opinions~\cite{baldassarri2008partisans}.

\section*{Acknowledgment}
This work was supported in part by the Army Research Laboratory (ARL) through the Cooperative Agreement (NS CTA) Number W911NF-09-2-0053, and by the Office of Naval Research (ONR) under Grant N00014-15-1-2640. The views and conclusions contained in this document are those of the authors and should not be interpreted as representing the official policies either expressed or implied of the Army Research Laboratory or the U.S. Government.

\section*{Supporting Material}
\subsection*{Derivation of the equilibrium points} 
The equilibrium points $x^*$ of the dynamical model defined by Eq. [2.1] satisfy the following equation:
\begin{equation}
    (1-x)x^a u_x = x(1-x)^a (1-u_x).
\end{equation}
Setting $w=x^*/(1-x^*)$ makes $w$ depending only on the equilibrium point $x^*$, and yields the following:
\begin{equation}\label{eq:a}
    a = \frac{\ln(1-u) - \ln u}{\ln w} + 1.
\end{equation}
Substituting $u=(1+z)/2$, yields: 
\begin{align}
    a &= \frac{\ln(1-z) - \ln(1+z) }{\ln w} + 1
\end{align}
which becomes a constant 1 when $z$ approaches $0$. Using the Taylor expansion of $\ln(1+z) = \sum_{j=1}^k -(-z)^j/j - O(z^{k+1})$ and 
$\ln(1-z) = \sum_{j=1}^k -z^j/j - O(z^{k+1})$, we get:
\begin{equation}
    a = \frac{-2z\sum_{j=0}^k z^{2j}/(2j+1) + O(z^{2k+2})}{\ln w} + 1 =  \frac{-2z\sum_{j=0}^k f(z,k) + O(z^{2k+2}) }{\ln w} + 1,
\end{equation}
where $f(z,k)=\sum_{j=0}^k\frac{z^{2j}}{2j+1}$. For $z \in (-1,1)$ for simplicity we can use the approximation with $k=0$: 
\begin{align}
    a \approx 1 - 2z\frac{1}{\ln w}
\end{align}
Given the particular $u$ and $a$ and substituting $w$ back with $x^*/(1-x^*)$ and $z$ with $u$, the equilibrium points $x^*$ of the dynamical system have the analytic\
 expression
\begin{align}
    x^* \approx \frac{1}{1+exp\left(2\frac{2u-1}{a-1}\right)}
\end{align}

To increase the precision of the approximation, we can use more terms in the Taylor expansion, for example
\begin{align} \label{eq:general}
    a \approx 1 - 2z\frac{f(z,k)}{\ln w}
\end{align}
decreases the relative error of approximation to the upper bound defined now as $z^{2k+2}/(2k+3)/(1-z^2)$. This approximation's relative error is small for quite 
large range, e.g., for $z\in (-0.6,0.6)$ and therefore $u\in (0.3, 0.8)$, it is less than 0.3\% for $k=2$ for the broader range  $z\in (-0.8,0.8)$ so for $u\in (0.1, 0.9)$, the precision of less than 0.15\% requires $k=5$. Applying the same transformation as before to Eq. [\ref{eq:general}], we get
\begin{align}
    x^* \approx \frac{1}{1+exp\left(2\frac{f(2u-1,k)}{a-1}\right)}.
\end{align}

When $a<1$, these equilibrium points are stable~\cite{abrams2011dynamics}. However, when $a>1$, there are some initial states that are tipping points 
from which the system nondeterministically converges to one of the two extreme equilibrium points, either to full consensus or to full polarization. In the context 
of political polarization, the existence of tipping points changes the stability of the polarization evolution. When $a > 1$, and with the system in the 
neighborhood of a tipping point, in response to even a small perturbation of the system state, the system may change the converges from one extreme equilibrium 
point to another. When $a<1$, the dynamical social system is more robust against the random perturbations because it always converges to the equilibrium point with some level of polarization.

%%%%%%%%%% Insert bibliography here %%%%%%%%%%%%%%

\end{document}